# Deterministic Identity Testing of Read-Once Algebraic Branching Programs

Maurice Jansen[*]    Youming Qiao[*]    Jayalal Sarma M.N.[*]

November 4, 2018

**Abstract**

In this paper we study polynomial identity testing of sums of $k$ read-once algebraic branching programs ($\Sigma_k$-RO-ABPs), generalizing the work of Shpilka and Volkovich [1, 2], who considered sums of $k$ read-once formulas ($\Sigma_k$-RO-formulas). We show that $\Sigma_k$-RO-ABPs are strictly more powerful than $\Sigma_k$-RO-formulas, for any $k \leq \lfloor n/2 \rfloor$, where $n$ is the number of variables. Nevertheless, as a starting observation, we show that the generator given in [2] for testing a single RO-formula also works against a single RO-ABP.

For the main technical part of this paper, we develop a property of polynomials called *alignment*. Using this property in conjunction with the *hardness of representation approach* of [1, 2], we obtain the following results for identity testing $\Sigma_k$-RO-ABPs, provided the underlying field has enough elements (more than $kn^4$ suffices):

1. Given free access to the RO-ABPs in the sum, we get a deterministic algorithm that runs in time $O(k^2 n^7 s) + n^{O(k)}$, where $s$ bounds the size of any largest RO-ABP given on the input. This implies we have a deterministic polynomial time algorithm for testing whether the sum of a constant number of RO-ABPs computes the zero polynomial.

2. Given black-box access to the RO-ABPs computing the *individual* polynomials in the sum, we get a deterministic algorithm that runs in time $k^2 n^{O(\log n)} + n^{O(k)}$.

3. Finally, given only black-box access to the polynomial computed by the sum of the $k$ RO-ABPs, we obtain an $n^{O(k + \log n)}$ time deterministic algorithm.

Items 1. and 3. above strengthen two main results of [2] (Theorems 2 and 3, respectively, for the case of non-preprocessed $\Sigma_k$-RO-formulas).

## 1 Introduction

In this paper we make contributions to the program of constructing increasingly more powerful pseudo-random generators useful against arithmetic circuits. As argued by Agrawal [3], this program is an approach towards resolving Valiant's Hypothesis, which states that the algebraic complexity classes VP and VNP are distinct.

Central to this program is the PIT problem: given an arithmetic circuit $C$ with input variables $x_1, x_2 \ldots x_n$ over a field $\mathbb{F}$, test if $C(x_1, x_2, \ldots, x_n)$ computes the zero polynomial in the ring

[*]Institute for Theoretical Computer Science, Tsinghua University, Beijing, P.R. China. Email: maurice.julien.jansen@gmail.com, jimmyqiao86@gmail.com, jayalal@tsinghua.edu.cn. This work was supported in part by the National Natural Science Foundation of China Grant 60553001, and the National Basic Research Program of China Grant 2007CB807900,2007CB807901.

$\mathbb{F}[x_1, x_2, \ldots x_n]$. This is a well-studied algorithmic problem with a long history and a variety of connections and applications. See [4] for a recent survey. Efficient randomized algorithms were proposed independently by Schwartz [5] and Zippel [6]. Obtaining a deterministic algorithm for the problem seemed surprisingly elusive.

It was originally Kabanets and Impagliazzo [7] who showed the strong connection between derandomizing PIT and proving circuit lower bounds. They showed that giving a deterministic polynomial time (even subexponential time) identity testing algorithm means either that NEXP $\not\subseteq$ P/$poly$, or that the permanent has no polynomial size arithmetic circuits. This was further strengthened in [3], where it was shown that giving a black-box derandomization of PIT implies that an explicit multilinear polynomial has no subexponential size arithmetic circuits.

Since the seminal work of [7], there has been a lot of attention and an impressive amount of progress in the area. Some of the special cases for which progress has been reported are: depth-2 arithmetic formulas [8, 9, 10], depth-3 and depth-4 arithmetic circuits with bounded top fan-in [11, 12, 13, 14, 15, 16], and non-commutative arithmetic formulas [17]. In a surprising result, Agrawal and Vinay [18] showed that the black-box derandomization of PIT for only depth-4 circuits is almost as hard as that for general arithmetic circuits.

Partly aimed at making progress towards an efficient deterministic PIT algorithm for multi-linear formulas, Shpilka and Volkovich [1, 2] studied the arithmetic read-once formula model. An arithmetic read-once formula is given by a tree whose nodes are taken from $\{+, \times\}$, and whose leaves are variables or field constants, subject to the restriction that each variables $x_i$ is allowed to appear at most once. In their work, efficient black-box deterministic PIT algorithms are given for $\Sigma_k$-RO-formulas, for "moderate" $k$.

We remark that due to a construction by Valiant [19], given a RO-formula $F$ of size $s$ computing $f$, one can express $f$ as a "read-once" determinantal expression $f = det(M)$, where $M$ is a $O(s)$-dimensional matrix, whose entries are variables or field elements. In this, each variable $x_i$ appears at most once in $M$. Identity testing read-once determinantal expressions, is an important special case of the PIT problem, as it is well-known that the bipartite perfect matching problem (BIPARTITE-PM) reduces to that form. Giving a black-box algorithm for testing such expressions has the potential of putting BIPARTITE-PM in NC, which is a prominent open problem in complexity theory regarding parallelizability [20, 21, 22, 23].

## 1.1 Results

We consider a generalization of the above mentioned RO-formulas, namely *read-once algebraic branching programs* (RO-ABP)[1]. An algebraic branching program (ABP) is a layered directed acyclic graph with two special vertices $s$ and $t$. Each edge is assigned a weight, which is an element of $X \cup \mathbb{F}$, where $X$ is a set of variables. For a path in the graph its weight is taken to be the product of the weight on its edges. The ABP itself computes a polynomial which is the sum of the weights of all paths from $s$ to $t$. The ABP is said to be *read-once* if each variable appears on at most one edge. A polynomial $f \in \mathbb{F}[X]$ is called a *RO-ABP-polynomial* if there exists a RO-ABP which computes $f$.

Due to [19], if $f$ can be computed by a RO-formula of size $s$, then $f$ can be computed by a RO-ABP of size $O(s)$. However, RO-ABPs are strictly more powerful than RO-formulas. Appendix A shows a RO-ABP computing $g = x_1x_2 + x_2x_3 + \cdots + x_{2n-1}x_{2n}$. Example 3.12 in [1] shows that $g$ can not be computed by a RO-formula, if $n \geq 2$. We remark that the RO-ABP model in not universal, e.g. for $n \geq 3$, $\prod_{1 \leq i < j \leq n} x_i x_j$, is not an RO-ABP-polynomial (See Appendix B). By [19], if $f$ is computable by a RO-ABP of size $s$, then we can write $f$ as a read-once determinantal expression $f = \det(M(x))$, where $M$ is a matrix of dimension $O(s)$.

---
[1]See Section 2 for a formal definition.

The results we will mention next make progress towards identity testing read-once determinantal expressions. This contributes to the program for separating VP and VNP mentioned in previous section (See e.g. [24] for a direct connection).

Our first result is to show that the Shpilka-Volkovich generator (SV-generator) used in [2] for identity testing RO-formulas also provides a test for RO-ABPs. This generator has also very recently been applied to identity testing multilinear depth 4 circuits with bounded top fan-in [16]. It is defined as follows:

Let $A = \{a_1, a_2, \ldots, a_n\} \subseteq \mathbb{F}$ be a set of size $n$. For every $i \in [n]$, let $u_i(w)$ be the $i$th Lagrange interpolation polynomial on $A$. Then $u_i(w)$ is a polynomial of degree $n-1$ satisfying that $u_i(a_j) = 1$ if $j = i$ and 0 otherwise. For every $i \in [n]$ and $k \geq 1$, define

$$G_k^i(y_1, y_2, \ldots, y_k, z_1, z_2, \ldots, z_k) = \sum_{j \in [k]} u_i(y_j) z_j.$$

and let $G_k(y_1, y_2, \ldots, y_k, z_1, z_2, \ldots, z_k) : \mathbb{F}^{2k} \to \mathbb{F}^n$, be defined by $G_k = (G_k^1, G_k^2, \ldots, G_k^n)$. We refer to the polynomial mapping $G_k$ as the $k$th-order SV-generator, or SV-generator for short. We have the following "Generator Lemma":

**Lemma 1.** *Let $f \in \mathbb{F}[X]$ be a nonzero RO-ABP-polynomial with $|var(f)| \leq 2^m$, for some $m \geq 0$. Then $f(G_{m+1}) \not\equiv 0$.*

To make further progress, we consider sums of $k$ RO-ABPs. We give an explicit *hitting-set* of size $n^{O(k + \log n)}$ for $\Sigma_k$-RO-ABPs. Namely we have the following theorem:

**Theorem 1.** *Let $\{f_i \in \mathbb{F}[X]\}_{i \in [k]}$ be a set of $k$ RO-ABPs. Let $f = \sum_{i \in [k]} f_i$. Provided $|\mathbb{F}| > kn^4$, we have that $f \equiv 0 \iff \forall a \in \mathcal{W}_{5k}^n + \mathcal{A}_k, f(a) = 0$, where $\mathcal{W}_k^n = \{y \in \{0,1\}^n \mid wt(y) \leq k\}$ and $\mathcal{A}_k = G_m(V^{2m})$ for the $m$th-order SV-generator with $m = \lceil \log n \rceil + 1$, and $V \subset \mathbb{F}$ is a arbitrary set of size $kn^4 + 1$.*

In the above for $V, W \subseteq F^n$, $V + W$ denotes the set $\{v + w : v \in V, w \in W\}$. By Theorem 1, we obtain the following black-box PIT for $\Sigma_k$-RO-ABPs:

**Theorem 2.** *Let $f = \sum_{i \in [k]} f_i$ be a sum of $k$ RO-ABP-polynomials in $n$ variables. Let $\mathbb{F}$ be a field with $|\mathbb{F}| > kn^4$. Given black-box access to $f$, it can be decided deterministically in time $n^{O(k + \log n)}$ whether $f \equiv 0$.*

This strengthens a main result of [2] (Theorem 3, for the non-preprocessed[2] case), which provides a deterministic $n^{O(k + \log n)}$ time PIT algorithm for $\Sigma_k$-RO-formulas. Namely, we prove a strict separation between $\Sigma_k$-RO-formula and $\Sigma_k$-RO-ABP, for $k \leq \lfloor n/2 \rfloor$. We show that

**Theorem 3.** $\prod_{i \in [2n], i \text{ is odd}} \prod_{j \in [2n], j \text{ is even}} x_i x_j$ *can not be written as a sum of $\lfloor n/2 \rfloor$ RO-formulas.*

The polynomial of Theorem 3 can be computed by a *single* RO-ABP of size $O(n^2)$ (see Section 3). In the non-black-box setting we will prove the following result:

---
[2] A generalization of our theorems to preprocessed $\Sigma_k$-RO-ABPs will not be pursued here.

**Theorem 4.** *Let $\{A_i\}_{i \in [k]}$ be a set of $k$ RO-ABPs in $n$ variables. Let $\mathbb{F}$ be a field with $|\mathbb{F}| > kn^2$. Given $\{A_i\}_{i \in [k]}$ on the input, it can be decided deterministically in time $O(k^2 n^7 s) + n^{O(k)}$ whether $\sum_{i \in [k]} f_i \equiv 0$, where $f_i$ is the RO-ABP-polynomial computed by $A_i$, for $i \in [k]$.*

Since the construction in [19] can be computed efficiently, this strengthens Theorem 2 in [2], for the case of non-preprocessed $\Sigma_k$-RO-formulas.

Finally, if black-box access is granted to the individual $f_i$'s, which we call the *semi-black-box* setting, we obtain the following result:

**Theorem 5.** *Let $\{f_i\}_{i \in [k]}$ be a set of $k$ RO-ABP-polynomials in $n$ variables. Let $\mathbb{F}$ be a field with $|\mathbb{F}| > kn^2$. Given black-box access to each individual $f_i$, it can decided deterministically in time $k^2 n^{O(\log n)} + n^{O(k)}$ whether $\sum_{i \in [k]} f_i \equiv 0$.*

## 1.2 Techniques for $\Sigma_k$-RO-ABP PIT

The results for $\Sigma_k$-RO-ABP PIT are obtained through the *hardness of representation* approach of [1, 2]. There the PIT algorithm is derived from a statement that $x_1 x_2 \ldots x_n$ cannot be expressed as a sum of $k \leq n/3$ RO-formula computable polynomials $\{f_i\}_{i \in [k]}$, if the polynomials $f_i$ satisfy some special property. We do not need to define this special property for the discussion here, except that we should name it: $\bar{0}$-justification.

Unfortunately, the property of $\bar{0}$-justification, does not work for the $\Sigma_k$-RO-ABP model. With some thought it can be seen that the monomial $x_1 x_2 \ldots x_n$ is expressible as the sum of three $\bar{0}$-justified RO-ABP-polynomials. Our main technical contribution is the development of a new "special property", called *alignment*, for which a hardness of representation theorem can still be proved, but which also can be satisfied simultaneously for a collection of RO-ABP-polynomials by means of an efficiently computable coordinate shift.

With regards to the latter, consider $f = f_1 + f_2 + \ldots + f_k$, where each $f_i$ is a RO-ABP-polynomial. Observe that $\forall v \in \mathbb{F}^n$, $f \equiv 0 \iff f(x_1 + v_1, x_2 + v_2, \ldots, x_n + v_n) \equiv 0$. With some technical work, we will establish a *sufficient* condition for alignment. With it we show that we can compute a coordinate shift $v$ such that all $f_i(x+v)$ are aligned. Such a shift $v$ is called a *simultaneous alignment*. In the case of having only black-box access to $f$, we will show we have a "small" set of candidates containing at least one simultaneous alignment. The PIT algorithms will follow from this.

The rest of this paper is organized as follows. Section 2 contains preliminaries. In Section 3 we compare $\Sigma_k$-RO-formulas and $\Sigma_k$-RO-ABPs. In Section 4 we prove Generator Lemma 1. In Section 5 we develop the tools regarding alignment. Then in Section 6 we show how to compute a simultaneous alignment. Section 7 contains the hardness of representation theorem for RO-ABPs. From these developments, we put the PIT algorithms together in Section 8.

## 2 Preliminaries

Let $X = \{x_1, x_2, \ldots, x_n\}$ be a set of variables and let $\mathbb{F}$ be a field. Let $\mathcal{W}_k^n = \{y \in \{0,1\}^n \mid wt(y) \leq k\}$, where $wt(y)$ counts the number of ones in $y$.

**Definition 1.** *(RO-ABPs) An algebraic branching program (ABP) is a 4-tuple $A = (G, w, s, t)$, where $G = (V, E)$ is an edge-labeled directed acyclic graph for which the vertex set $V$ can be parti-* tioned into levels $L_0, L_1, \ldots, L_d$, where $L_0 = s$ and $L_d = t$. Vertices $s$ and $t$ are called the source and sink of $B$, respectively. Edges may only go between consecutive levels $L_i$ and $L_{i+1}$.

The label function $w : E \to X \cup \mathbb{F}$ assigns variables or field constants to the edges of $G$. For a path $p$ in $G$, we extend the weight function by $w(p) = \prod_{e \in p} w(e)$. Let $P_{i,j}$ denote the collection of all directed paths $p$ from $i$ to $j$ in $G$. The program $A$ computes the polynomial $\hat{A} := \sum_{p \in P_{s,t}} w(p)$. The size of $A$ is defined to be $|V|$.

An ABP is said to be *read-once* if $|w^{-1}(x_i)| \leq 1$, for each $x_i \in X$. That is, every variable is read at most once by the program. A polynomial $f \in \mathbb{F}[X]$ is called a *RO-ABP-polynomial*, if there exists a RO-ABP which computes $f$. We use the following notation: for $x_i$ present on arc $(v, w)$ in a RO-ABP $A$: $begin(x_i) = v$ and $end(x_i) = w$. We let $source(A)$ and $sink(A)$ stand for the source and sink of $A$. For any nodes $v, w$ in $A$, we denote the subprogram with source $v$ and sink $w$ by $A_{v,w}$. A *layer* of a RO-ABP $A$ is any subgraph induced by two consecutive levels $L_i$ and $L_{i+1}$ in $A$. We will assume RO-ABPs are in the form given by the following straightforwardly proven lemma:

**Lemma 2.** *If $f \in \mathbb{F}[X]$ is a RO-ABP-polynomial, then $f$ can be computed by a RO-ABP $A$, where every layer contains at most one variable-labeled edge.*

Let $f$ be a polynomial in the ring $\mathbb{F}[X]$. For $\alpha \in \mathbb{F}$, $f|_{x_i = \alpha}$ denotes the polynomial $f(x_1, x_2, \ldots x_{i-1}, \alpha, x_{i+1}, \ldots, x_n)$. Extending this to sets of variables, for a subset $I \subseteq [n]$ and an assignment $a \in \mathbb{F}^n$, $f|_{x_I = a_I}$ is the the polynomial resulting from setting the variable $x_i$ to $a_i$ in $f$ for every $i \in I$. This is not to be confused with the following notation: for $S \subseteq \mathbb{F}^n$, we will write $f_{|S} \equiv 0$ to denote that $\forall a \in S, f(a) = 0$.

The following two notions are taken from [2]. We say that a polynomial $f$ *depends on a variable* $x_i$ if there exists an $a \in \mathbb{F}^n$ and $b \in \mathbb{F}$, such that $f(a_1, a_2, a_{i-1}, a_i, a_{i+1}, \ldots, a_n) \neq f(a_1, a_2, a_{i-1}, b, a_{i+1}, \ldots, a_n)$. The set of variables $x_i$ that $f$ depends on is denoted by $Var(f)$. For a polynomial $f \in \mathbb{F}[X]$, the *partial derivative with respect to* $x_i$, denoted by $\frac{\partial f}{\partial x_i}$, is defined as $f|_{x_i=1} - f|_{x_i=0}$. We will freely use the properties listed for this notion in [2]. For example, a multilinear polynomial $f$ depends on $x_i$ if and only if $\frac{\partial f}{\partial x_i} \not\equiv 0$. In addition, $\frac{\partial f}{\partial x_i}$ does not depend on $x_i$. Partial derivatives commute, which we express by saying that $\frac{\partial^2 f}{\partial x_i x_j} = \frac{\partial^2 f}{\partial x_j x_i}$. Setting values to variables commutes with taking partial derivatives in the following way: $\forall i \neq j$, $\frac{\partial f}{\partial x_i}|_{x_j=a} = \frac{\partial (f|_{x_j=a})}{\partial x_i}$.

**Lemma 3.** *Let $f \in \mathbb{F}[X]$ be a RO-ABP-polynomial, then $\frac{\partial f}{\partial x_i}$ is a RO-ABP-polynomial.*

*Proof.* Let $p = |var(f)|$. In case $p = 0$ it is trivial. Assume $p > 0$. If $x_i \notin var(f)$, then $\frac{\partial f}{\partial x_i} \equiv 0$, in which case the property trivially holds. Now suppose $x_i \in var(f)$. Hence $x_i$ must appear somewhere in $A$. Say $x_i$ is on the arc $(v_1, w_1)$ from level $L_j$ to $L_{j+1}$, where $L_j = \{v_1, v_2, \ldots, v_{m_1}\}$ and $L_{j+1} = \{w_1, w_2, \ldots, w_{m_2}\}$, for certain $j, m_1, m_2$. We can write

$$f = \sum_{a \in [m_1]} \sum_{b \in [m_2]} f_{s, v_a} w(v_a, w_b) f_{w_b, t}, \qquad (1)$$

where for any nodes $p$ and $q$ in $A$, $f_{p,q}$ is the polynomial computed by subprogram $A_{p,q}$. Then

$$\begin{aligned}
\frac{\partial f}{\partial x_i} &= f_{|x_i=1} - f_{|x_i=0} \\
&= \sum_{a\in[m_1]}\sum_{b\in[m_2]} f_{s,v_a} w(v_a,w_b)_{|x_i=1} f_{w_b,t} - \sum_{a\in[m_1]}\sum_{b\in[m_2]} f_{s,v_a} w(v_a,w_b)_{|x_i=0} f_{w_b,t} \\
&= \sum_{a\in[m_1]}\sum_{b\in[m_2]} f_{s,v_a} \left(w(v_a,w_b)_{|x_i=1} - w(v_a,w_b)_{|x_i=0}\right) f_{w_b,t} \\
&= f_{s,v_1} f_{w_1,t}.
\end{aligned}$$

Hence we obtain a valid RO-ABP computing $\frac{\partial f}{\partial x_i}$ from $A$ by setting the label of the wire $(v_1, w_1)$ to 1, and removing all other wires between layers $L_j$ and $L_{j+1}$. □

The proof of the above lemma provides the insight that a RO-ABP computing $\frac{\partial f}{\partial x_i}$ can be obtained from a RO-ABP computing $f$, by setting $x_i = 1$ and removing all other edges in the layer containing $x_i$. This fact will be used at several places in the paper. Finally, observe the following simple-but-useful factor-lemma:

**Lemma 4.** *If $f \in \mathbb{F}[X]$ is a RO-ABP-polynomial such that $f \not\equiv 0$ and $f = g \cdot (\beta x_i - \alpha)$, then $g$ is a RO-ABP-polynomial.*

*Proof.* This follows from the fact that for every $\gamma$ with $\beta\gamma - \alpha \neq 0$, $g = \frac{1}{\beta\gamma - \alpha} \cdot f_{|x_i=\gamma}$. □

## 2.1 Combinatorial Nullstellensatz and a Lemma by Gauss

**Lemma 5** (Lemma 2.1 in [25]). *Let $f \in \mathbb{F}[X]$ be a nonzero polynomial such that the degree of $f$ in $x_i$ is bounded by $r_i$, and let $S_i \subseteq \mathbb{F}$ be of size at least $r_i + 1$, for all $i \in [n]$. Then there exists $(s_1, s_2, \ldots, s_n) \in S_1 \times S_2 \times \ldots \times S_n$ with $f(s_1, s_2, \ldots, s_n) \neq 0$.*

**Lemma 6.** (Gauss) *Let $P \in \mathbb{F}[X, y]$ be a nonzero polynomial, and let $g \in \mathbb{F}[X]$ be such that $P|_{y=g(x)} \equiv 0$. Then $y - g(x)$ is an irreducible factor of $P$ in the ring $\mathbb{F}[X]$.*

## 3 Separation of RO-ABP and $\Sigma_{\lfloor n/2 \rfloor}$-RO-formulas

For $n \geq 2$, let $f_n$ be defined as

$$f_n(x_1, x_2, \ldots, x_{2n-1}, x_{2n}) = \prod_{i \in [2n], i \text{ is odd}} \prod_{j \in [2n], j \text{ is even}} x_i x_j.$$

**Proposition 1.** *$f_n$ can be computed by an RO-ABP of size $O(n^2)$.*

*Proof.* The RO-ABP is shown in Figure 1. Note that between the $(n+1)$th level and the $(n+2)$th level there is an $n$ by $n$ complete bipartite graph. □

**Proposition 2.** *A polynomial $p(x_1, x_2, \ldots, x_n)$ that contains three terms of form $\alpha x_i x_j + \beta x_j x_k + \gamma x_k x_l$, where $i, j, k, l \in [n]$ are pairwise different, and $\alpha, \beta, \gamma \in \mathbb{F}$ are nonzero, can not be computed by a RO-formula, for $n \geq 4$.*

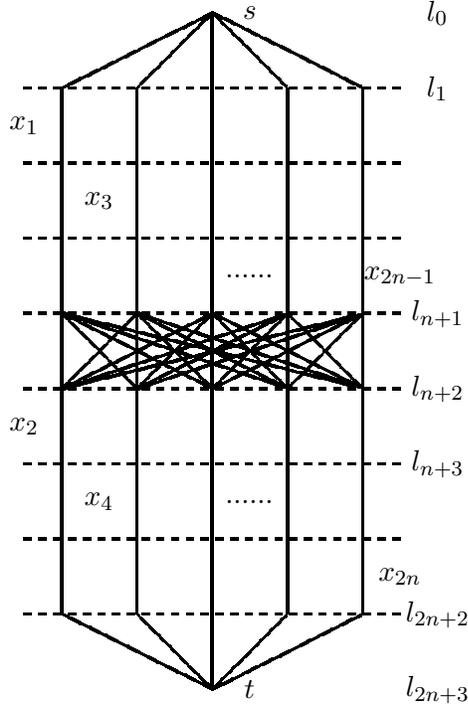

Figure 1: A RO-ABP computing $f_n$.

*Proof.* For the purpose of contradiction, suppose there is a RO-formula $F$ computing $p$. Setting all $x_m = 0$, for $m \in [n] \setminus \{i, j, k, l\}$, would result in an RO-formula $F'$ computing $p'(x_i, x_j, x_k, x_l) = \alpha x_i x_j + \beta x_j x_k + \gamma x_k x_l + a x_i + b x_j + c x_k + d x_l + e$. However, $p'$ can not be computed by an RO-formula. One argues this in a similar manner as for $x_1 x_2 + x_2 x_3 + x_3 x_4$ (See example 3.12 in [1]). □

Consider the complete bipartite graph $G_n = (V_n, E_n)$ for $f_n$, called the graph associated with $f_n$, shown in Figure 2. Every edge represents a term in $f_n$. The term $x_i x_j + x_j x_k + x_k x_l$ can be viewed as a length-3 path in $G_n$.

**Proposition 3.** *Let $n \geq 2$. In $G_n$, for an edge set $S \subseteq E_n$ with $|S| \geq 2n - 1$, $S$ must contain a length-3 path.*

*Proof.* We just need to prove that for $G_n$, the maximum "length-3 path free" edge set is of size at most $2(n - 1)$. This is proved by induction on $n$. For $n = 2$, it is easy to see that it holds. Suppose for $n < l$ the claim holds. Then for $n = l$, for any length-3 path free edge set $S$, consider the following two cases:

1. If there exists an edge $e = (u, v) \in S$, for which $u$ or $v$ has no other outgoing edges, let $S' = S \setminus \{e\}$. $S'$ is a length-3 path free set in $G_{l-1}$. By induction, $|S'| \leq 2(l - 2)$. Thus $S$ has at most $1 + 2(l - 2) < 2(l - 1)$ edges.

2. Otherwise, partition the vertices adjacent to edges in $S$ into two sets $V_1$ and $V_2$, where $V_1$ contains all vertices of degree one, and $V_2$ contains all vertices of degree larger than one.

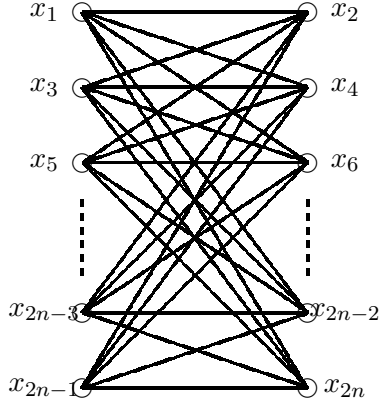

Figure 2: The bipartite graph $G_n$ for $f_n$.

It is noted that since no length-3 paths exist, we have that $|S| = |V_1|$. If $|V_2| \geq 2$, then $|V_1| \leq 2l - 2 = 2(l - 1)$, since there are at most $2l$ vertices adjacent to edges in $S$. In case $|V_2| = 1$, then $S$ is a star, i.e. a single vertex $u$ connected to a collection of vertices $v_1, v_2, \ldots, v_k$. Then $k \leq l$ and $|S| = k \leq l \leq 2(l - 1)$, for $l \geq 2$.

□

**Theorem 6.** *$f_n$ can not be represented as a sum of $\lfloor n/2 \rfloor$ RO-formulas.*

*Proof.* For the purpose of contradiction, suppose $f_n$ can be represented as a sum of $\lfloor n/2 \rfloor$ RO-formula-polynomials $q_1, q_2, \ldots, q_{\lfloor n/2 \rfloor}$. Let $G_n = (V_n, E_n)$ be the graph associated with $f_n$. For any $q_i$, let $S_i \subseteq E_n$ be the set of edges representing the terms appearing in $q_i$ of the form $x_a x_b$, where $a \in [2n]$ is even, and $b \in [2n]$ is odd. Note that since $f$ has $n^2$ many terms, some $q_i$ should have $|S_i| \geq 2n$. Then by Claim 3, $S_i$ contains a length-3 path. Therefore $\alpha x_i x_j + \beta x_j x_k + \gamma x_k x_l$ appears in $q_i$, for distinct $i, j, k$ and nonzero constants $\alpha, \beta, \gamma \in \mathbb{F}$. Due to Claim 2, $q_i$ can not be computed by a RO-formula, which is a contradiction. □

## 4 Proof of Generator Lemma 1

Let $p = |Var(f)|$. The proof proceeds by induction on $p$. The bases $p = 0$ and $p = 1$ trivially hold.

Suppose $p > 1$. Hence $m \geq 1$. Consider arbitrary RO-ABP $A$ computing $f$. Let $s$ and $t$ be the source and sink of $A$, respectively. Wlog. assume that only the $p$ variables in $Var(f)$ are present in $A$, and assume $A$ satisfies the condition yielded by Lemma 2. Observe that for some variable $x_i$ there are at most $p/2$ variables in layers before the layer containing $x_i$, and at most $p/2$ variables in layers after. (If $p$ is odd it splits $((p-1)/2), (p-1)/2)$ if $p$ is even it splits $(p/2 - 1, p/2)$).

Say $x_i$ is on the arc $(v_1, w_1)$ from layer $L_j$ to $L_{j+1}$, where $L_j = \{v_1, v_2, \ldots, v_{m_1}\}$ and $L_j = \{v_1, v_2, \ldots, v_{m_2}\}$, for certain $j, m_1, m_2$. We can write

$$f = \sum_{a=1}^{m_1} f_{s,v_a} f_{v_a,t}, \tag{2}$$

where for any nodes $p$ and $q$ in $A$, $f_{p,q}$ is the polynomial computed by subprogram of $A_{p,q}$. Consider $f' = f(G_m^1, \ldots, G_m^{i-1}, x_i, G_m^{i+1}, \ldots, G_m^n)$.

**Claim 1.** *Write $f' = x_i \cdot \frac{\partial f}{\partial x_i}(G_m^1,, \ldots, G_m^{i-1}, G_m^{i+1}, \ldots, G_m^n) + f(G_m^1,, \ldots, G_m^{i-1}, 0, G_m^{i+1}, \ldots, G_m^n)$. Then $\frac{\partial f}{\partial x_i}(G_m^1,, \ldots, G_m^{i-1}, G_m^{i+1}, \ldots, G_m^n) \not\equiv 0$.*

*Proof.* Since $f$ depends on $x_i$ and $f$ is multilinear, $\frac{\partial f}{\partial x_i} \not\equiv 0$. Let $f'' = \frac{\partial f}{\partial x_i}$. We will show that $f''(G_m) \not\equiv 0$. Observe that in the r.h.s. of (2) only $f_{v_1,t}$ depends on $x_i$. This implies that $f'' = \frac{\partial f_{v_1,t}}{\partial x_i} \cdot f_{s,v_1}$. Observe that $|Var(f_{s,v_1})|$ and $|Var(\frac{\partial f_{v_1,t}}{\partial x_i})|$ are both at most $p/2$. Since $f'' \not\equiv 0$, both $f_{s,v_1}$ and $\frac{\partial f_{v_1,t}}{\partial x_i}$ are not identically zero. Certainly $f_{s,v_1}$ can be computed by a RO-ABP. By Lemma 3, we know also $\frac{\partial f_{v_1,t}}{\partial x_i}$ can be computed by a RO-ABP. As $p/2 < p$, the induction hypothesis applies. Since $p/2 \leq 2^{m-1}$, it yields that $f_{s,v_1}(G_m) \not\equiv 0$ and $\frac{\partial f_{v_1,t}}{\partial x_i}(G_m) \not\equiv 0$. Therefore $f''(G_m) \not\equiv 0$. This proves the claim. □

Recall the set $A = \{a_1, \ldots, a_n\}$ used for the construction of the SV-generator. By Observation 5.2 in [2], $f(G_{m+1})_{|y_{m+1}=a_i} = f'_{|x_i = G_m^i + z_{m+1}}$. Since $z_{m+1}$ does not appear in $G_m^j$ for any $j$, we get by Claim 1 that $f(G_{m+1})_{|y_{m+1}=a_i} \not\equiv 0$. Hence $f(G_{m+1}) \not\equiv 0$. □

## 5 X-Aligned RO-ABP-polynomials

The following lemma leads up to our central definition:

**Lemma 7.** . *For all $i \in [k]$, Let $f \in \mathbb{F}[X]$ be a RO-ABP-polynomial with $|Var(f)| \geq 3$. Then for any $x_i \in Var(f)$, there exist distinct $x_j, x_k \in X \setminus \{x_i\}$ such that $\frac{\partial^2 f}{\partial x_j \partial x_k} = g \cdot (\beta x_i - \alpha)$, where $g$ is a RO-ABP-polynomial that does not depend on $x_i$, and $\alpha, \beta \in \mathbb{F}$.*

*Proof.* Let $A$ be a RO-ABP computing $f$. Wlog. assume all variables in $X$ appear in $A$. By Lemma 2 assume wlog. that $A$ has at most one variable per layer. Let $x_{r_1}, x_{r_2}, \ldots, x_{r_n}$ be the variables in $X$ as they appear layer-by-layer, when going from the source to the sink of $A$. Consider an arbitrary $x_i \in Var(f)$. First, we handle the case that $i = r_m$, for some $1 < m < n$.

Let $j = r_{m-1}$ and $k = r_{m+1}$. So $x_j$ and $x_k$ are the variables right before and right after $x_i$ in $A$, respectively. Assume that $x_j$ and $x_k$ label the edges $(u, v)$ and $(m, n)$ respectively. Then $\frac{\partial^2 f}{\partial x_j \partial x_k} = f_{s,u} f_{v,m} f_{n,t}$, where $f_{s,u} f_{v,m}$, and $f_{n,t}$ are computed by the subprograms $A_{s,u}, A_{v,m}$, and $A_{n,t}$, respectively. Observe that $f_{v,m}$ is of form $\beta x_i - \alpha$, for $\alpha, \beta \in \mathbb{F}$. Take $g = f_{s,u} f_{v,m}$, which is easily seen to be RO-ABP-computable by putting $A_{s,u}$ and $A_{v,m}$ in series, or by appealing to Lemmas 3 and 4.

The special case where $i = r_1$ ($i = r_n$), i.e. $x_i$ is the first (last) variable in $A$, is handled similarly as above, by choosing $x_k \in X \setminus \{x_i, x_j\}$ arbitrarily and appealing to Lemma 3. □

In the above lemma we have no guarantee the $\alpha$ is nonzero, in case $\beta \neq 0$. We would like to consider polynomials which are in general position in this regard. We make the following definition:

**Definition 2.** *Let $S \subseteq X$. Every RO-ABP-polynomial $f \in \mathbb{F}[X]$ with $|Var(f)| \leq 2$ is $X$-pre-aligned on $S$. A RO-ABP-polynomial $f \in \mathbb{F}[X]$ with $|Var(f)| > 2$ is $X$-pre-aligned on $S$, if the following condition is satisfied:*

1. for every $x_i \in S$, there exist distinct $x_j, x_k \in X \setminus \{x_i\}$ such that $\frac{\partial^2 f}{\partial x_j \partial x_k} = g \cdot (\beta x_i - \alpha)$, where $g$ is a RO-ABP-polynomial that does not depend on $x_i$, and $\alpha, \beta \in F$ satisfy that $\alpha = 0 \Rightarrow \beta = 0$.

If $f$ is $X$-pre-aligned on $Var(f)$, we simply say that $f$ is $X$-pre-aligned.

For the $X$-pre-alignment property to hold recursively w.r.t. setting variables to zero, is a particularly desirable property of a RO-ABP-polynomial to have, as we will see. We make the following inductive definition:

**Definition 3.** *Every RO-ABP-polynomial $f \in \mathbb{F}[X]$ with $|Var(f)| \leq 2$ is $X$-aligned. A RO-ABP-polynomial $f \in \mathbb{F}[X]$ with $|Var(f)| > 2$ is $X$-aligned, if the following conditions are satisfied:*

1. *$f$ is $X$-pre-aligned, and*
2. *for every $x_i \in Var(f)$, $f_{|x_i=0}$ is $X \setminus \{x_i\}$-aligned.*

Next we prove some of the needed properties of our notion, starting with the following easily verified statement:

**Proposition 4.** *If $f \in \mathbb{F}[X]$ is $X$-pre-aligned, then $\forall \mu \in \mathbb{F}$, $\mu \cdot f$ is $X$-pre-aligned. The same statement holds with aligned instead of pre-aligned.*

The notion of $X$-pre-alignment is well-behaved w.r.t. taking partial derivatives. This will be crucial for obtaining the Hardness of Representation Theorem 8. We have the following lemma:

**Lemma 8.** *For any RO-ABP-polynomial $f \in \mathbb{F}[X]$ and any $x_r \in X$, the following hold:*

1. *If $f$ is $X$-pre-aligned, then $\frac{\partial f}{\partial x_r}$ is $(X \setminus \{x_r\})$-pre-aligned.*
2. *If $f$ is $X$-aligned, then $\frac{\partial f}{\partial x_r}$ is $(X \setminus \{x_r\})$-aligned.*

*Proof.* We first show that Item 1 holds. Let $f' = \frac{\partial f}{\partial x_r}$ and $X' = X \setminus \{x_r\}$. By Lemma 3, we know that $f'$ is a RO-ABP-polynomial. Assume that $|Var(f')| \geq 3$, since otherwise the statement holds trivially. Consider arbitrary $x_i \in Var(f')$. Then $x_i \in Var(f)$, so there exist distinct $x_j$ and $x_k$ in $X \setminus \{x_i\}$, such that $\frac{\partial^2 f}{\partial x_j \partial x_k} = g \cdot (\beta x_i - \alpha)$, where $g$ is a RO-ABP-polynomial that does not depend on $x_i$, and $\alpha = 0 \Rightarrow \beta = 0$. Consider the following two cases:

**Case I:** $r \notin \{j, k\}$.

Hence $x_j, x_k \in X' \setminus \{x_i\}$. We have that $\frac{\partial^2 f'}{\partial x_j \partial x_k} = \frac{\partial^3 f}{\partial x_j \partial x_k \partial x_r} = \frac{\partial g}{\partial x_r} \cdot (\beta x_i - \alpha)$. By Lemma 3, $\frac{\partial g}{\partial x_r}$ is a RO-ABP-polynomial, and it clearly does not depend on $x_i$, so we conclude that $f'$ is $X'$-pre-aligned on $\{x_i\}$.

**Case II:** $r \in \{j, k\}$.

Wlog. assume $r = j$. Then $x_k \in X' \setminus \{x_i\}$. Since $|Var(f')| \geq 3$, there must be at least one more variable $x_l$ in $Var(f')$ distinct from each of $x_k$ and $x_i$. Then $x_l \in X' \setminus \{x_i\}$. We have that $\frac{\partial f'}{\partial x_k} = g \cdot (\beta x_i - \alpha)$. Hence $\frac{\partial^2 f'}{\partial x_k \partial x_l} = \frac{\partial g}{\partial x_l} \cdot (\beta x_i - \alpha)$. We again conclude $f'$ is $X'$-pre-aligned on $\{x_i\}$.

Since in the above, $x_i$ was taken arbitrarily from $Var(f')$, we conclude $f'$ is $X'$-pre-aligned.

Item 2 is proved by induction on $|X|$. The base case is when $|X| \leq 3$. Then $|Var(f')| \leq 2$, and hence $f'$ is $X'$-aligned. Now suppose $|X| > 3$. Assume $|Var(f')| > 2$, since otherwise it is trivial. By Item 1, we know $f'$ is $X'$-pre-aligned. Consider an arbitrary $x_i \in Var(f')$. Then $x_i \in Var(f)$.

We have that $f'_{|x_i=0} = \left(\frac{\partial f}{\partial x_r}\right)_{x_i=0} = \frac{\partial f_{|x_i=0}}{\partial x_r}$. Since $f_{|x_i=0}$ is $(X\backslash\{x_i\})$-aligned, we can apply the induction hypothesis to conclude that $\frac{\partial f_{|x_i=0}}{\partial x_r}$ is $(X\backslash\{x_i\})\backslash\{x_r\} = (X'\backslash\{x_i\})$-aligned. □

## 5.1 A Workable Sufficient Condition

Next we establish a sufficient condition, so for a given RO-ABP-polynomial $f$ we can make $f(x_1 + v_1, x_2 + v_2, \ldots, x_n + v_n)$ $X$-aligned, by means of computing some shift $v \in \mathbb{F}^n$. For this, let us call a polynomial $f \in \mathbb{F}[X]$ *decent*, if for all $x_a, x_b \in Var(f)$ with $\frac{\partial^2 f}{\partial x_a \partial x_b} \not\equiv 0$, it holds that the monomial $x_a x_b$ appears in $f$ with a nonzero constant coefficient.

**Lemma 9.** *A RO-ABP-polynomial $f \in \mathbb{F}[X]$ is $X$-aligned, if $|Var(f)| \leq 2$, or else for any $I \subseteq Var(f)$ with $|I| \leq |Var(f)| - 3$, $f_{|x_I=0}$ is decent.*

*Proof.* We use induction on $|Var(f)|$. For the base case $|Var(f)| \leq 2$ it is trivial. Now assume $|Var(f)| > 2$. Take $I = \emptyset$. Then we get that for any $x_a, x_b \in Var(f)$, if $\frac{\partial^2 f}{\partial x_a \partial x_b} \not\equiv 0$ then the monomial $x_a x_b$ appears in $f$ with a nonzero constant coefficient.

Let us first establish that $f$ is $X$-pre-aligned. Consider an arbitrary $x_i \in Var(f)$. By Lemma 7, there exist distinct $x_j, x_k \in X \backslash \{x_i\}$ such that

$$\frac{\partial^2 f}{\partial x_j \partial x_k} = g \cdot (\beta x_i - \alpha), \tag{3}$$

where $g$ is a RO-ABP-polynomial that does not depend on $x_i$, and $\alpha, \beta \in F$.

If $\beta = 0$, then $f$ is $X$-pre-aligned on $\{x_i\}$, so suppose $\beta \neq 0$. If (3) is identically zero, then we know $g \equiv 0$, so $\frac{\partial^2 f}{\partial x_j \partial x_k} = g \cdot (\beta x_i - \alpha')$, for any arbitrary $\alpha' \neq 0$. If (3) is not identically zero, then we know $x_j x_k$ is in $f$, which implies that $\alpha \neq 0$. We conclude that $f$ is $X$-pre-aligned on $\{x_i\}$.

In the above, we find that $f$ is $X$-pre-aligned on $\{x_i\}$ in any of the considered cases. Since $x_i$ was arbitrarily taken from $Var(f)$, we conclude that $f$ is $X$-pre-aligned.

Next, we show Condition 2 of Definition 3 holds. Consider $f' := f_{|x_i=0}$, for an arbitrary $x_i \in Var(f)$. We want to establish that the sufficient condition of Lemma 9 holds for $f' \in \mathbb{F}[X \backslash \{x_i\}]$, since then we can by apply the induction hypothesis and conclude that $f'$ is $(X \backslash \{x_i\})$-aligned.

If $|Var(f')| \leq 2$ the sufficient condition of the Lemma 9 clearly holds for $f'$. Otherwise, consider $I' \subseteq Var(f')$ of size at most $|Var(f')| - 3$. Let $I = I' \cup \{x_i\}$. Then $|I| \leq |Var(f)| - 3$. Now consider $x_a, x_b \in Var(f'_{x_{I'}=0}) = Var(f_{x_I=0})$. Suppose $\frac{\partial^2 f'_{|x_{I'}=0}}{\partial x_a \partial x_b} \not\equiv 0$. Since the latter equals $\frac{\partial^2 f_{|x_I=0}}{\partial x_a \partial x_b} \not\equiv 0$, we know that $x_a x_b$ appears with a nonzero constant coefficient in $f_{|x_I=0}$. This implies $x_a x_b$ appears with a nonzero constant coefficient in $f_{|x_{I'}=0}$. Hence $f'_{x_{I'}=0}$ is decent.

We conclude the sufficient condition of the Lemma 9 holds for $f' \in \mathbb{F}[X \backslash \{x_i\}]$. Hence by the induction hypothesis we conclude that $f'$ is $(X \backslash \{x_i\})$-aligned. □

**Lemma 10.** *Any decent RO-ABP-polynomial $f \in \mathbb{F}[X]$ is $X$-aligned.*

*Proof.* We show that the condition of Lemma 9 is satisfied. If $|Var(f)| \leq 2$ this is clear. Otherwise, consider arbitrary $I \subseteq Var(f)$ with $|I| \leq |Var(f)| - 3$. Let $x_a, x_b \in Var(f_{|x_I=0})$, be such that $\frac{\partial^2 f_{|x_I=0}}{\partial x_a \partial x_b} \not\equiv 0$. We have that $x_a, x_b \in Var(f)$, and it must be that $\frac{\partial^2 f}{\partial x_a \partial x_b} \not\equiv 0$, since $\frac{\partial^2 f_{|x_I=0}}{\partial x_a \partial x_b} = \left(\frac{\partial^2 f}{\partial x_a \partial x_b}\right)_{|x_I=0}$. Hence $x_a x_b$ is in $f$. This implies that $x_a x_b$ is in $f_{|x_I=0}$. □

## 5.2 Nearly Unique Nonalignment

In addition to the above, we crucially need the following "Nearly Unique Nonalignment Lemma".

**Lemma 11.** *Let $f \in \mathbb{F}[X]$ be an $X$-pre-aligned RO-ABP-polynomial for which $\frac{\partial^2 f}{\partial x_p \partial x_q} \not\equiv 0$, for any distinct $x_p, x_q \in X$. Then there are at most two $\gamma \in \mathbb{F}$ such that $f_{|x_n=\gamma}$ is not $(X\backslash\{x_n\})$-pre-aligned.*

Before giving the proof, we need a lemma.

**Lemma 12.** *Let $f \in \mathbb{F}[X]$ be a RO-ABP-polynomial with $|Var(f)| \geq 3$ that is $X$-pre-aligned on $S$, for some $S \subseteq Var(f)$. Assume that for any distinct $x_p, x_q \in X$, $\frac{\partial^2 f}{\partial x_p \partial x_q} \not\equiv 0$. In any RO-ABP $A$ computing $f$, for any $x_i \in S$,*

1. *if there exists a non-constant layer with variable $x_a$ right before the $x_i$-layer, and there exists a non-constant layer with variable $x_b$ right after the $x_i$-layer, then*

$$\frac{\partial^2 f}{\partial x_a \partial x_b} = g \cdot (\beta x_i - \alpha),$$

*where $g$ is a RO-ABP-polynomial that does not depend on $x_i$, and $\alpha, \beta \in F$ satisfy that $\alpha = 0 \Rightarrow \beta = 0$. Furthermore, $-\alpha$ equals the sum of weights of all paths from $end(x_a)$ to $begin(x_b)$ that do not go over $x_i$.*

*Proof.* Consider $x_i \in S$. Since $f$ is $X$-pre-aligned on $S$, we know there exist distinct $x_j, x_k \in X\backslash\{x_i\}$ with $\frac{\partial^2 f}{\partial x_j \partial x_k} = h \cdot (\beta' x_i - \alpha')$, where $h$ is a RO-ABP-polynomial that does not depend on $x_i$, and $\alpha', \beta' \in F$ satisfy that $\alpha' = 0 \Rightarrow \beta' = 0$. Since $\frac{\partial^2 f}{\partial x_j \partial x_k} \not\equiv 0$, it must be that $\alpha' \neq 0$.

**Case I:** In $A$, the $x_i$-layer lies in between the $x_j$-layer and $x_k$ layer.

Wlog assume the $x_i$ layer lies before the $x_k$-layer and after the $x_j$-layer (according to the order of the DAG underlying $A$). Write $\frac{\partial^2 f}{\partial x_j \partial x_k} = p_1 p_2 \cdot (q_1 q_2 x_i + q_3)$, where

- $p_1$ is the sum of weights over all paths in $A$ from $source(A)$ to $begin(x_j)$, and $p_2$ is the sum of weights over all paths in $A$ from $end(x_k)$ to $sink(A)$.

- $q_3$ is the sum of weights over all paths from $end(x_j)$ to $begin(x_k)$ that bypass the $x_i$-edge, $q_1$ is the sum of weights over all paths from $end(x_j)$ to $begin(x_i)$, and $q_2$ is the sum of weights over all paths from $end(x_i)$ to $begin(x_k)$.

Now we have that $p_1 p_2 \cdot (q_1 q_2 x_i + q_3) = h \cdot (\beta' x_i - \alpha')$. Since both $p_1 p_2$ and $h$ do not depend on $x_i$, it must be that $(\beta' x_i - \alpha') \mid (q_1 q_2 x_i + q_3)$. Note that $\beta'$ cannot equal 0, since then one of $q_1, q_2$ would be zero. The latter implies that $\frac{\partial^2 f}{\partial x_i \partial x_j} \equiv 0$ or $\frac{\partial^2 f}{\partial x_i \partial x_k} \equiv 0$, which is a contradiction. Since $\beta' \neq 0$, we can conclude that $q_3 = \mu q_1 q_2$ for some $\mu \in \mathbb{F}, \mu \neq 0$. Now we need the following claim:

**Claim 2.** *Given an RO-ABP $A$ computing $f(x_1, \ldots, x_n)$, if for any distinct $x_p, x_q \in X$, $\frac{\partial^2 f}{\partial x_p \partial x_q} \not\equiv 0$, then $\prod_{i \in [n]} x_i$ appears in $f$. Furthermore, for two variables $x_i$ and $x_j$, if $x_i$ is before $x_j$ in $A$, if we let $S$ be the set of variables in between $x_i$ and $x_j$, then $\prod_{x_m \in S} x_m$ is a term in the polynomial $\hat{A}(end(x_i), begin(x_j))$.*

*Proof.* Suppose the variable layers in $A$ are arranged according to the permutation $\phi: [n] \to [n]$, that is, $x_{\phi(i)}$ labels the $i$th variable layer. Then we that

1. $\hat{A}(s, begin(x_{\phi(1)})) \not\equiv 0$ (Since otherwise $\frac{\partial^2 f}{\partial x_{\phi(1)} \partial x_{\phi(2)}} \equiv 0$),

2. Similarly $\hat{A}(end(x_{\phi(n)}), t) \not\equiv 0$, and

3. For $i \in [n-1]$, $\hat{A}(begin(x_{\phi(i)}), end(x_{\phi(i+1)})) \not\equiv 0$ (Since otherwise $\frac{\partial^2 f}{\partial x_{\phi(i)} \partial x_{\phi(i+1)}} \equiv 0$).

The coefficient of $\prod_{i \in [n]} x_i$ is just

$$\hat{A}(s, begin(x_{\phi(1)})) \cdot \hat{A}(end(x_{\phi(n)}), t) \prod_{i \in [n-1]} \hat{A}(begin(x_{\phi(i)}), end(x_{\phi(i+1)})),$$

and hence $\prod_{i \in [n]} x_i$ appears in $f$. A similar argument yields the statement for $\hat{A}(end(x_i), begin(x_j))$. □

As in the proof of Lemma 7, write $\frac{\partial^2 f}{\partial x_a \partial x_b} = g \cdot (\beta x_i - \alpha)$, where $g$ is a RO-ABP-polynomial that does not depend on $x_i$, and $-\alpha$ equals the sum of weights over all paths from $end(x_a)$ to $begin(x_b)$ not going over $x_i$. We have three cases:

1. Neither $x_j$ nor $x_k$ is the most adjacent variable to $x_i$ in $A$. By above claim, $x_a$ appears in a monomial of $q_1$, and $x_b$ appears in a monomial $q_2$. Hence, there is a monomial in $q_1 q_2$ with $x_a x_b$. As $q_3 = \mu q_1 q_2$, for $\mu \neq 0$, the same can be said for $q_3$. But this implies $\alpha \neq 0$, as the coefficient of $x_a x_b$ is $-\alpha \cdot \hat{A}(end(x_j), begin(x_a)) \hat{A}(end(x_b), begin(x_k))$.

2. $x_j$ is not the most adjacent variable to $x_i$ in $A$, but $x_k = x_b$. Then similarly $q_1 q_2$ has a monomial with $x_a$ in it, and therefore the same holds for $q_3$. Therefore $\alpha \neq 0$, as the coefficient of $x_a$ in $q_3$ is $-\alpha \cdot \hat{A}(end(x_j), begin(x_a))$.

3. $x_j = x_a$, but $x_k$ is not the most adjacent variable to $x_i$ in $A$. This is argued similarly as the second item.

This concludes the argument for this case.

**Case II:** In $A$, the $x_i$-layer lies before the $x_j$-layer and $x_k$-layer.

Wlog. assume that the $x_j$ layer lies before the $x_k$ layer. Similarly as in Case I, we write $\frac{\partial^2 f}{\partial x_j \partial x_k} = p_1 p_2 \cdot (q_1 q_2 x_i + q_3)$, but where now we have that

- $p_1 = \hat{A}_{end(x_j), begin(x_k)}$, and $p_2 = \hat{A}_{end(x_k), sink(A)}$,

- $q_1 = \hat{A}_{source(A), begin(x_i)}$,

- $q_2 = \hat{A}_{end(x_i), begin(x_j)}$,

- $q_3 = A[\hat{x_i = 0}]_{source(A), begin(x_j)}$.

Then $p_1 p_2 \cdot (q_1 q_2 x_i + q_3) = h \cdot (\beta' x_i - \alpha')$. Since both $p_1 p_2$ and $h$ do not depend on $x_i$, it must be that $(\beta' x_i - \alpha') \mid (q_1 q_2 x_i + q_3)$. Similarly as before, we get $q_3 = \mu q_1 q_2$ for some $\mu \in \mathbb{F}$, $\mu \neq 0$.

The rest of the proof is similar to Case I. One argues that 1) when $x_j \neq x_b$, $q_1 q_2$ contains a monomial with $x_a x_b$. To make $x_a x_b$ appear in a monomial $q_3$ we need $\alpha \neq 0$, and 2) when $x_j = x_b$, $q_1 q_2$ contains a monomial with $x_a$, and to make $x_a$ appear in a monomial of $q_3$, we need $\alpha \neq 0$.

**Case III:** In $A$, the $x_i$-layer lies after the $x_j$-layer and $x_k$-layer.
This case is symmetrical to Case II. □

We also need the following proposition:

**Proposition 5.** *Let $f \in \mathbb{F}[X]$ be a RO-ABP-polynomial with $|Var(f)| \geq 3$, and let $S \subseteq Var(f)$. Then $f$ is $X$-pre-aligned on $S$ if and only if $f' := (x_{n+1} + 1)f$ is $X \cup \{x_{n+1}\}$-pre-aligned on $S$.*

*Proof.* Let $X' = X \cup \{x_{n+1}\}$. It is easy to see that assuming $f$ is $X$-pre-aligned on $S$, we have that $f$ is $X'$-pre-aligned on $S$.

Conversely, assume $f'$ is $X'$-pre-aligned on $S$. Let $x_i \in S$. Then there exist $x_j, x_k \in X' \setminus \{x_i\}$, such that $\frac{\partial^2 f'}{\partial x_j \partial x_k} = g(\beta x_i + \alpha)$, where $g$ is a RO-ABP-polynomial that does not depend on $x_i$, and $\alpha = 0$ implies $\beta = 0$. If $x_{n+1} \notin \{x_j, x_k\}$, then $\frac{\partial^2 f'}{\partial x_j \partial x_k} = \frac{\partial^2 f}{\partial x_j \partial x_k}(x_{n+1} + 1)$. Setting $x_{n+1} = 0$, we have that $\frac{\partial^2 f}{\partial x_j \partial x_k} = (g_{|x_{n+1}=0})(\beta x_i + \alpha)$. So we get the required $X$-pre-alignment of $f$ on $\{x_i\}$. Otherwise, say wlog. $x_j = x_{n+1}$. We have that $\frac{\partial f}{\partial x_k} = \frac{\partial^2 f'}{\partial x_{n+1} \partial x_k} = g(\beta x_i + \alpha)$. One easily obtains the required $X$-pre-alignment of $f$ on $\{x_i\}$, by taking one more $\partial x_l$, for some variable $x_l \in X \setminus \{x_i, x_k\}$, and then using Lemma 3. □

We are now ready to give the proof of Lemma 11.

## 5.3 Proof

We prove the lemma by induction on $|X|$. For the base case we take $|X| \leq 3$, in which case the statement clearly holds. Now suppose $|X| > 3$. Let $f' = f_{|x_n=\gamma}$, for some $\gamma$. Let $X' = X \setminus \{x_n\}$. Suppose $f'$ is not $X'$-pre-aligned. Hence $|Var(f')| \geq 3$. We want to show this can happen for at most one $\gamma$.

Consider an arbitrary RO-ABP $A$ computing $f$. Let $f_e = f(x_{n+1}+1)(x_{n+2}+1)(x_{n+3}+1)(x_{n+4}+1)$. Let $X_e := X \cup \{x_{n+1}, x_{n+2}, x_{n+3}, x_{n+4}\}$. By Proposition 5, $f_e$ is $X_e$-pre-aligned on $Var(f)$. Let $f'_e := (f_e)_{|x_n=\gamma}$ and $X'_e := X' \cup \{x_{n+1}, x_{n+2}, x_{n+3}, x_{n+4}\}$. Note that $f'_e = f'(x_{n+1}+1)(x_{n+2}+1)(x_{n+3}+1)(x_{n+4}+1)$. So also by Proposition 5, $f'_e$ is not $X'_e$-pre-aligned on $Var(f')$ if and only if $f'$ is not $X'$-pre-aligned on $Var(f')$. We will show the former happens for at most one $\gamma$. So let us assume that $f'_e$ is not $X'_e$-pre-aligned on $Var(f')$. We can easily obtain a RO-ABP $A_e$ from $A$, which computes $f_e$. In this, we make sure $x_{n+1}$ and $x_{n+2}$ are the first and second variable in $A_e$, and $x_{n+3}$ and $x_{n+4}$ are the fore-last and last variable in $A_e$. For each $x_i \in Var(f')$, let $x_{j_i}$ be the variable right after $x_i$ in $A^e$, and let $x_{k_i}$ be the variable before $x_i$ in $A_e$. Note that we have made sure these always exist in $A_e$. Since $f_e$ is $X_e$-pre-aligned on $Var(f)$, by Lemma 12, $\frac{\partial^2 f_e}{\partial x_{j_i} \partial x_{k_i}} = g \cdot (\beta_i x_i - \alpha_i)$, where $g$ is a RO-ABP-polynomial that does not depend on $x_i$, and $\alpha_i = 0 \Rightarrow \beta_i = 0$. Furthermore, we have that $\alpha_i$ is the sum of weights of all paths from $end(x_{k_i})$ to $begin(x_n)$, which do not go over $x_i$ in $A_e$. Consider the following two cases:

**Case I:** $n \notin \{j_i, k_i\}$, for any $x_i \in Var(f')$.

Then for any $i$, $\frac{\partial^2 f'_e}{\partial x_{j_i} \partial x_{k_i}} = (g_i)_{|x_n=\gamma} \cdot (\beta_i x_i - \alpha_i)$, which contradicts the assumption that $f'_e$ is not $X'_e$-pre-aligned on $Var(f')$.

**Case II:** $n \in \{j_i, k_i\}$, for some $x_i \in Var(f')$.

By symmetry we can assume wlog. that $j_i = n$ (the case $k_i = n$ is handled similarly). Since $\frac{\partial^2 f}{\partial x_{j_i} \partial x_{k_i}} \neq 0$, and $\alpha_i = 0$ implies $\beta_i = 0$, We have that $\alpha_i \neq 0$.

We know that in $A_e$ there still exists a variables layer, say with variables $x_l$, right after the $x_{j_i}$-layer. Let $b_i = begin(x_i), e_i = end(x_i), b_n = begin(x_n)$, and $e_n = end(x_n)$. Let $s = end(x_{k_i})$ and $t = begin(x_l)$. Then write:

$$\frac{\partial^2 f_e}{\partial x_l \partial x_{k_i}} = p_1 p_2 (c_{s,b_i} c_{e_i,b_n} c_{e_n,t} x_i x_n + c_{s,b_i} c_{e_i,t} x_i + c_{s,b_n} c_{e_n,t} x_n + c_{s,t}),$$

where in the above each constant $c_{v,w}$ is the sum of weights over all paths from $v$ to $w$ going over constant labeled edges only. Note that $c_{s,b_n} = \alpha_i \neq 0$. Furthermore, $p_1$ is the sum of weights of all paths from $source(A_e)$ to $begin(x_{k_i})$, and $p_2$ is the sum of weights over all paths from $end(x_l)$ to $sink(A_e)$. Then

$$\frac{\partial^2 f'_e}{\partial x_l \partial x_{k_i}} = p_1 p_2 ((c_{s,b_i} c_{e_i,b_n} c_{e_n,t} \gamma + c_{s,b_i} c_{e_i,t}) x_i + c_{s,b_n} c_{e_n,t} \gamma + c_{s,t}),$$

We have that $f'_e$ can only not be $X'_e$-pre-aligned on $\{x_i\}$ if $c_{s,b_n} c_{e_n,t} \gamma + c_{s,t} = 0$. This can happen for more than one $\gamma$ only if $c_{s,b_n} c_{e_n,t} = 0$. Since $c_{s,b_n} \neq 0$, this happens only if $c_{e_n,t} = 0$, but the latter implies that $\frac{\partial^2 f_e}{\partial x_l \partial x_n} \equiv 0$, which in turn implies that $\frac{\partial^2 f}{\partial x_l \partial x_n} \equiv 0$, which is a contradiction.

Finally, putting together from what we observed from the above two cases, note that, Case II can apply at most twice for a variable $x_i \in Var(f')$. Namely, possibly once for the variable right before $x_n$, and possibly once for the variable after $x_n$. We conclude the lemma holds. □

**Corollary 1.** *Suppose $|\mathbb{F}| > 3$. Let $h, g \in \mathbb{F}[X]$ be RO-ABP-polynomials such that $h = g \cdot (\beta x_n - \alpha)$, for $\beta \in \mathbb{F} \backslash \{0\}$. If $h$ is $X$-pre-aligned, then $g$ is $(X \backslash \{x_n\})$-pre-aligned.*

*Proof.* If we set $x_n$ to any value $\gamma \neq \alpha/\beta$, we get that $h_{|x_n = \gamma}$ is a nonzero constant multiple of $g$. By Lemma 11, there are at most two $\gamma$ such that $h_{|x_n = \gamma}$ is not $(X \backslash \{x_n\})$-pre-aligned. Now use Proposition 4 to conclude that $g$ is $(X \backslash \{x_n\})$-pre-aligned. □

## 6 Simultaneous Alignment of RO-ABP-polynomials

**Definition 4.** *A simultaneous $X$-alignment for a set of RO-ABP-polynomials $\{f_i \in \mathbb{F}[X]\}_{i \in [k]}$ is a vector $v \in \mathbb{F}^n$ such that $f_i(x_1 + v_1, x_2 + v_2, \ldots, x_n + v_n)$ is $X$-aligned for every $i \in [k]$.*

We present an algorithm for finding a simultaneous $X$-alignment for a set of RO-ABP-polynomials. We assume that we have a polynomial identity testing algorithm $\text{PIT}_{\text{RO-ABP}}$ for testing a single RO-ABP. We prove a corollary of Lemma 10 first.

**Corollary 2.** *Let $\{f_i\}_{i \in [k]}$ be a set of RO-ABP-polynomials in $\mathbb{F}[X]$. Then $v \in \mathbb{F}^n$ is a simultaneous $X$-alignment for $\{f_i\}_{i \in [k]}$, if it is a simultaneous nonzero for $\{\frac{\partial^2 f_i}{\partial x_a \partial x_b} \mid \frac{\partial^2 f_i}{\partial x_a \partial x_b} \not\equiv 0\}_{i \in [k], a, b \in [n]}$.*

*Proof.* Consider $\{f'_i = f_i(x_1 + v_1, x_2 + v_2, \ldots, x_n + v_n)\}_{i \in [k]}$. Due to Lemma 10, we only need to show that for every $i$, for every $x_a, x_b \in Var(f_i)$, if $\frac{\partial^2 f'_i}{\partial x_a \partial x_b} \not\equiv 0$ then the monomial $x_a x_b$ appears in $f'_i$ with a nonzero constant coefficient. Observe that the monomial $x_a x_b$ appears in $f'_i$ with a nonzero constant coefficient $\iff \frac{\partial^2 f'_i}{\partial x_a \partial x_b}(\bar{0}) \neq 0$. The latter holds, as $\frac{\partial^2 f'_i}{\partial x_a \partial x_b}(\bar{0}) = \frac{\partial^2 f_i}{\partial x_a \partial x_b}(v) \neq 0$. □

Now the argument is similar as for Lemma 4.3 in [2], but with first order partial derivatives replaced by second order ones. This yields the following theorem:

**Theorem 7.** *Let $\mathbb{F}$ be a field with $|\mathbb{F}| > kn^2$. There exists an algorithm for finding a simultaneous X-alignment for a set of RO-ABP polynomials $\{f_i \in \mathbb{F}[X]\}_{i \in [k]}$. The algorithm makes oracle calls to the procedure $\mathrm{PIT}_{\text{RO-ABP}}$. The $f_i$s are only accessed through this subroutine. The running-time of the algorithm is $O(k^2 n^5 \cdot t)$, where $t$ is an upper bound on the time needed for any subroutine call to $\mathrm{PIT}_{\text{RO-ABP}}$.*

*Proof.* We assume that we have a polynomial identity testing algorithm $\mathrm{PIT}_{\text{RO-ABP}}$ for testing a single RO-ABP, such that $\mathrm{PIT}_{\text{RO-ABP}}$ outputs $True$ if $f \equiv 0$ and $False$ otherwise. We have the following algorithm:

---
**Algorithm 1** Alignment Finding.
---
**Input:** A set of RO-ABP-polynomials $\{f_i \in \mathbb{F}[X]\}_{i \in [k]}$.
**Output:** A simultaneous alignment $v$ for $\{f_i\}_{i \in [k]}$.
**Oracle:** PIT algorithm $\mathrm{PIT}_{\text{RO-ABP}}$.

1: $L = \emptyset$
2: **for all** $f_i$ and $(x_a, x_b)$, $a, b \in [n]$, $a \neq b$ **do**
3:    If $\mathrm{PIT}_{\text{RO-ABP}}(\frac{\partial^2 f_i}{\partial x_a \partial x_b}) = False$, add it to $L$
4: **end for**
5: **for all** $j \in [n]$ **do**
6:    Find $c$ such that for every $g \in L$, $\mathrm{PIT}_{\text{RO-ABP}}(g \mid_{x_j=c}) = False$
7:    $v_j \leftarrow c$
8:    For every $g \in L$, $g \leftarrow g \mid_{x_j=c}$
9: **end for**
10: **return** $v$

---

We first make two remarks, which pertain to applying Algorithm 1 in the setting where we only have black-box access to each $f_i$. Consider the set $L$ the algorithm constructs with the execution of the first **for**-loop. Since we only have black-box access to $f_i$, the given pseudocode is intended to mean $L$ is constructed symbolically. Having black-box access to $f_i$ is enough to have black-box access to any element of $L$. Namely, by Lemma 3, $f' := \frac{\partial^2 f_i}{\partial x_a \partial x_b}$ is a RO-ABP. Note that black-box access to $f_i$ is sufficient for being able to compute $f'(a)$ for any $a \in \mathbb{F}^n$. This is all the black-box RO-ABP algorithm needs to decide whether $f' \equiv 0$.

Similarly, on line 8 the substitution is not actually carried out, but done symbolically. So it is just remembered that $x_j$ is set to $c$. For example, suppose that up to some point in the execution the algorithm it has set $x_i = c_i$, for $i \in [m]$. Then on line 6, for evaluating $\mathrm{PIT}_{\text{RO-ABP}}(g \mid_{x_j=c})$, the black-box algorithm is granted access to a RO-ABP in $n - m$ variables $g(c_1, c_2, \ldots, c_m, x_{m+1}, \ldots, x_n)$. The queries it makes can be answered with only black-box access to $g$.

Now, by Corollary 2 it suffices to find a common nonzero of the set $L$. First however, we need to explain how to find $c$ such that $g \mid_{x_j=c} \not\equiv 0$. Let $V \subset \mathbb{F}$ with $|V| = kn^2 + 1$ be given. We claim $V$ always includes a good value. This is because we have at most $kn^2$ multilinear polynomials in $L$, and for a specific one there is at most one bad value, due to Lemma 6. The algorithm can simply try all elements in $V$ to get the required $c$. The correctness of the algorithm is now evident, from the observation that it simply maintains the invariant that all $g \in L$ are not identically zero.

The running time of the algorithm is as follows: for line 2 we need $O(kn^2)$ calls to $\text{PIT}_{\text{RO-ABP}}$. For line 7 we need $O(n \cdot (kn^2+1) \cdot (kn^2)) = O(k^2n^5)$ calls to $\text{PIT}_{\text{RO-ABP}}$. Thus the total running time of the algorithm is $O(k^2n^5 \cdot t)$, where $t$ is an upper bound on the time needed for any subroutine call to $\text{PIT}_{\text{RO-ABP}}$. □

By Lemma 1 and using Lemma 5, $\text{PIT}_{\text{RO-ABP}}$ can be implemented in the black-box setting to run in time $n^{O(\log n)}$, where $n$ is the number of variables of the input RO-ABP-polynomial. In the non-black-box setting, as is show in Appendix C, $\text{PIT}_{\text{RO-ABP}}$ can be implemented to run in time $O(n^2 s)$, when given an RO-ABP over $n$ variables of size $s$. This yields the following two corollaries:

**Corollary 3.** *Provided $|\mathbb{F}| > kn^2$, there exists an non-black-box algorithm for finding a simultaneous $X$-alignment for a set $\{f_i \in \mathbb{F}[X]\}_{i \in [k]}$, where $f_i$ is computed by a RO-ABP $A_i$, for $i \in [k]$. The algorithm receives $\{A_i\}_{i \in [k]}$ on the input, and it runs in time $O(k^2n^7 s)$, where $s$ is an upper bound on the size of any $A_i$.*

**Corollary 4.** *Provided $|\mathbb{F}| > kn^2$, there exists a black-box algorithm for finding a simultaneous $X$-alignment for a set of RO-ABP-polynomials $\{f_i \in \mathbb{F}[X]\}_{i \in [k]}$. The algorithm queries individual $f_i$s, and runs in time $k^2 n^{O(\log n)}$.*

## 6.1 Simultaneous Alignment Hitting Set

Here we present a black-box algorithm to find a candidate set $\mathcal{A}_k$ of size $(kn)^{O(\log n)}$, which is guaranteed to contain a simultaneous $X$-alignment for any set of $k$ RO-ABP-polynomials $\{f_i \in \mathbb{F}[X]\}_{i \in [k]}$.

**Lemma 13.** *Let $\mathbb{F}$ be a field with $|\mathbb{F}| > kn^4$, and let $V \subseteq \mathbb{F}$ with $|V| = kn^4 + 1$ be given. Let $\{f_i\}_{i \in [k]}$ be a set of RO-ABP-polynomials in $\mathbb{F}[X]$. Let $G_m : \mathbb{F}^{2m} \to \mathbb{F}^n$ be the mth-order SV-generator with $m = \lceil \log n \rceil + 1$. Then $\mathcal{A}_k := G_m(V^{2m})$ contains a simultaneous $X$-alignment for $\{f_i\}_{i \in [k]}$.*

*Proof.* let $L = \{\frac{\partial^2 f_i}{\partial x_a \partial x_b} \mid \frac{\partial^2 f_i}{\partial x_a \partial x_b} \not\equiv 0\}_{i \in [k], a, b \in [n]}$. Let $P(x_1, \ldots, x_n) = \prod_{g \in L} g(x_1, \ldots, x_n)$. By Lemma 3, each $g \in L$ is a RO-ABP-polynomial. Hence by Lemma 1, for $m = \lceil \log n \rceil + 1$, the SV-generator $(G_m^1, G_m^2, \ldots, G_m^n)$, satisfies that $g(G_m^1, G_m^2, \ldots, G_m^n) \not\equiv 0$, for all $g \in L$. So $P(G_m^1, G_m^2, \ldots, G_m^n) \not\equiv 0$.

Note that there are $2m$ variables in $P(G_m^1, \ldots, G_m^n)$, and the degree of every variable is bounded by $kn^2 \cdot n^2 = kn^4$. Thus by Lemma 5, $\exists a \in V^{2m}, P(G_m^1(a), \ldots, G_m^n(a)) \neq 0$. Hence $\mathcal{A}_k = G_n(V^{2m})$ is ensured to contain a nonzero of $P$. Any nonzero of $P$ is a simultaneous nonzero of all $g \in L$. By Corollary 2, $\mathcal{A}_k$ contains a simultaneous $X$-alignment for $\{f_i\}_{i \in [k]}$. □

## 7 A Hardness of Representation Theorem for RO-ABPs

The following theorem is an adaption of Theorem 6.1 in [2] to the notion of $X$-pre-alignment. One notable difference in the proof is that for the main case separation, we distinguish between whether there are 3rd-order partial derivatives vanishing or not (rather than 2nd-order partial as in [2]).

**Theorem 8.** *Assume $|\mathbb{F}| > 3$. Let $P_n = \prod_{i \in [n]} x_i$. If $\{f_i \in \mathbb{F}[X]\}_{i \in [k]}$ is a set of $k$ $X$-pre-aligned RO-ABP-polynomials for which $P_n = \sum_{i \in [k]} f_i$, then $n < 7k$.*

*Proof.* The proof proceeds by induction on $k$. For the base case $k = 1$, since $f_1 = P_n$, and $f_1$ is $X$-pre-aligned, it must be that $n \leq 2$. Namely, if $n > 2$, then for $x_i \in Var(P_n)$, whatever distinct $x_j, x_k \in X \setminus \{x_i\}$ we select, $\frac{\partial^2 f_1}{\partial x_j \partial x_k} = x_i \cdot \prod_{x_r \in X \setminus \{x_i, x_j, x_k\}}$. This cannot be of the form $g \cdot (\beta x_i + \alpha)$ with $g$ being an RO-ABP not depending on $x_i$, and $\alpha = 0 \Rightarrow \beta = 0$, as Definition 2 requires. Namely, since $g$ does not depend on $x_i$, it must be that $\beta \neq 0$. Hence $\alpha \neq 0$, and thus $g \cdot (\beta x_i + \alpha)$ is not homogeneous. Since $x_i \cdot \prod_{x_r \in X \setminus \{x_i, x_j, x_k\}}$ is homogeneous, this is a contradiction.

Now assume $k > 1$. Suppose we can write $P_n = \sum_{i \in [k]} f_i$. For purpose of contradiction, assume that $n \geq 7k$. Hence $n \geq 14$.

**Case I:** $\exists$ distinct $p, q, r \in [n]$ and $s \in [k]$, such that $\frac{\partial^3 f_s}{\partial x_p \partial x_q \partial x_r} \equiv 0$.

Wlog. assume that $p = n-2, q = n-1, r = n$ and $s = k$. Then $\sum_{i \in [k-1]} \frac{\partial^3 f_i}{\partial x_{n-2} \partial x_{n-1} \partial x_n} = P_{n-3}$.

By Lemma 8, all of the terms $\frac{\partial^3 f_i}{\partial x_{n-2} \partial x_{n-1} \partial x_n}$ are $(X \setminus \{x_{n-2}, x_{n-1}, x_n\})$-pre-aligned. By induction, it must be that $n - 3 < 5(k-1)$. Hence $n < 5k - 2$, which is a contradiction.

**Case II:** $\nexists$ distinct $p, q, r \in [n]$ and $s \in [k]$, such that $\frac{\partial^3 f_s}{\partial x_p \partial x_q \partial x_r} \equiv 0$.

We know $\forall i, |Var(f_i)| \geq 3$. Since $f_i$ is $X$-pre-aligned, there exist distinct $x_{j_i}, x_{k_i} \in X \setminus \{x_i\}$ such that $\frac{\partial^2 f}{\partial x_{j_i} \partial x_{k_i}} = g_i \cdot (\beta_i x_n - \alpha_i)$, where $g_i$ is a RO-ABP-polynomial that does not depend on $x_i$, and $\alpha_i = 0 \Rightarrow \beta_i = 0$. Note that in this case, $g_i \not\equiv 0$, since otherwise a second order partial vanishes. Hence both $j_i$ and $k_i$ are certainly not equal to $x_n$. It must be that $\beta_i \neq 0$, since otherwise $\frac{\partial^3 f}{\partial x_{j_i} \partial x_{k_i} \partial x_n} \equiv 0$. Hence also $\alpha_i \neq 0$.

**Claim 3.** *Any $g_i$ is $(X \setminus \{x_{j_i}, x_{k_i}, x_n\})$-pre-aligned.*

*Proof.* Assume that $|Var(g_i)| \geq 3$, since otherwise the claim is trivial. Let $h = g_i \cdot (\beta_i x_n - \alpha_i)$. By Lemma 8, $h$ is $(X \setminus \{x_{j_i}, x_{k_i}\})$-pre-aligned. Since $\beta_i \neq 0$, applying Corollary 1 yields that $g_i$ is $(X \setminus \{x_{j_i}, x_{k_i}, x_n\})$-pre-aligned. □

Now, let $A = \{\frac{\alpha_i}{\beta_i} : i \in [k]\}$. Define for $\gamma \in A$, $E_\gamma = \{i \in [k] : \gamma = \frac{\alpha_i}{\beta_i}\}$ and $B_\gamma = \{i \in [k] : \gamma \neq \frac{\alpha_i}{\beta_i}$ and $(f_i)_{|x_n=\gamma}$ is not $(X \setminus \{x_n\})$-pre-aligned$\}$. Note that $\sum_{\gamma \in A} |E_\gamma| = k$. By Nearly Unique Nonalignment Lemma 11, $\sum_{\gamma \in A} |B_\gamma| \leq 2k$. Hence there exists $\gamma_0 \in A$ such that $|B_{\gamma_0}| \leq 2|E_{\gamma_0}|$. Let $I = E_{\gamma_0} \cup B_{\gamma_0}$, and let $J = \{j_i : i \in I\} \cup \{k_i : i \in I\}$. We have that $2 \leq |J| \leq 2|I| \leq 6|E_{\gamma_0}|$. Observe that $x_n \notin J$. Define for any $i$, $f'_i = \partial_J f_i$. We have the following three properties:

1. Each $f'_i$ is an $(X \setminus J)$-pre-aligned RO-ABP-polynomial, due to Lemma 8.

2. For every $i \in I$, $f'_i = (\beta_i x_n - \alpha_i) h_i$, where $h_i$ is a RO-ABP-polynomial. Namely, since $j_i, k_i \in J$, $f'_i = \partial_{J \setminus \{j_i, k_i\}} [g_i (\beta_i x_n - \alpha_i)] = (\beta_i x_n - \alpha_i) \cdot \partial_{J \setminus \{j_i, k_i\}} g_i$.

3. In the above, each $h_i$ is an $(X \setminus (J \cup \{x_n\}))$-pre-aligned RO-ABP-polynomial. Namely, by Claim 3, $g_i$ is $(X \setminus \{x_{j_i}, x_{k_i}, x_n\})$-pre-aligned. Hence, using Lemma 8, we get that $h_i$ is an $(X \setminus (J \cup \{x_n\}))$-pre-aligned RO-ABP-polynomial.

For any $i$, define $f''_i = (f'_i)_{|x_n=\gamma_0}$. Then we have the following three properties:

1. $\forall i \in E_{\gamma_0}$, $f''_i \equiv 0$.

2. $\forall i \in B_{\gamma_0}$, $f''_i = (\beta_i \gamma_0 - \alpha_i) h_i$, so $f''_i$ is an $(X \setminus (J \cup \{x_n\}))$-pre-aligned RO-ABP-polynomial, due to Proposition 4.

3. For every $i \in [k] \setminus I$, $(f_i)_{|x_n = \gamma_0}$ is $X \setminus \{x_n\}$-pre-aligned. Since $n \notin J$, $f_i'' = (f_i')_{|x_n = \gamma_0} = \partial_J[f_{|x_n=\gamma_0}]$. So by Lemma 8, $f_i''$ is an $(X \setminus (J \cup \{x_n\}))$-pre-aligned RO-ABP-polynomial.

Wlog. assume that $J = \{\tilde{n}+1, \tilde{n}+2, \ldots, n-2, n-1\}$. Then $|J| = n - 1 - \tilde{n}$. Then $\sum_{i \in [k]} f_i'' = (\partial_J P_n)_{|x_n=\gamma_0} = \gamma_0 \cdot P_{\tilde{n}}$. Let $\tilde{X} = \{x_1, \ldots, x_{\tilde{n}}\}$. We have found a representation of $P_{\tilde{n}}$ as a sum of $\tilde{k}$ $\tilde{X}$-pre-aligned RO-ABP-polynomials, where $7\tilde{k} \leq 7(k - |E_{\gamma_0}|) \leq n - 7|E_{\gamma_0}| = n - 1 - 6|E_{\gamma_0}| + 1 - |E_{\gamma_0}| \leq \tilde{n} + 1 - |E_{\gamma_0}| \leq \tilde{n}$. This contradicts the induction hypothesis, and hence $n < 7k$. □

## 8 A Vanishing Theorem and the PIT Algorithms

The following theorem is analogous to Theorem 6.4 in [2].

**Theorem 9.** *Suppose $|\mathbb{F}| > 3$. Let $\{f_i \in \mathbb{F}[X]\}_{i \in [k]}$ be a set of $k$ $X$-aligned RO-ABPs. Let $f = \sum_{i \in [k]} f_i$. Then $f \equiv 0 \iff f|_{\mathcal{W}_{7k}^n} \equiv 0$.*

We need to argue only the "⇐"-direction. Assume that $f|_{\mathcal{W}_{7k}^n} \equiv 0$.

We use induction on the number of variables $n$. The base case is when $n < 7k$. In this case it follows from Lemma 5 that $f \equiv 0$.

For the induction case assume $n \geq 7k$. We restrict one variable at a time. Consider a variable $x_\ell$, for $\ell \in [n]$. Consider a restriction of the polynomials $f_i$'s and $f$ to the subspace $x_\ell = 0$.

By condition 2 in the definition of *aligned*, each of the restricted polynomials $f_i' = f_i|_{x_\ell=0}$ are $(X \setminus \{x_\ell\})$-aligned. Let $f' = \sum_{i=1}^k f_i'$. Clearly, $f'|_{\mathcal{W}_{7k}^{n-1}} = f'|_{\mathcal{W}_{7k}^n} \equiv 0$. Thus from the induction hypothesis, $f' = f|_{x_\ell=0} \equiv 0$, which implies that $x_\ell$ divides $f$. Since $\ell$ was arbitrarily chosen, this implies that $P_n = \prod_{i=1}^k x_i$ divides $f$. But since $f$ is multilinear, this gives $f = c \cdot P_n$ where $c$ is a constant and $P_n = \prod_{i \in [n]} x_i$.

Thus $c \cdot P_n$ is the sum of $k$ RO-ABPs which are also $X$-aligned (and therefore certainly $X$-pre-aligned). Since $n \geq 7k$, by Theorem 8, we can conclude that $c = 0$. Hence $f \equiv 0$. □

Now we are ready to give the identity testing algorithms for $\Sigma_k$-RO-ABP-polynomials given by $\{f_i \in \mathbb{F}[X]\}_{i \in [k]}$. The algorithm is simple. We use the fact that that $\forall v \in \mathbb{F}^n$, $f \equiv 0 \iff f(x_1 + v_1, x_2 + v_2, \ldots, x_n + v_n) \equiv 0$. Assuming that we have some common alignment $v$ for $\{f_i\}_{i \in [k]}$, we know that each $f_i(x_1 + v_1, x_2 + v_2, \ldots, x_n + v_n)$ is $X$-aligned. In this case, Theorem 9 is applicable, and it suffices to test if the polynomial evaluates to zero on the set $\mathcal{W}_{7k}^n$. Based on the three approaches to get a common alignment, the algorithms are as follows:

1. (*Non-black-box setting*) By Corollary 3, we obtain a simultaneous alignment in time $O(k^2 n^7 s)$. Then it takes $n^{O(k)}$ to test all points in $\mathcal{W}_{7k}^n$, so the running-time is $O(k^2 n^7 s) + n^{O(k)}$. This proves Theorem 4. In this case we need $|\mathbb{F}| > kn^2$.

2. (*Semi-black-box setting*) By Corollary 4, we obtain a simultaneous alignment in time $k^2 n^{O(\log n)}$. Then it takes $n^{O(k)}$ to test all points in $\mathcal{W}_{7k}^n$, so the running-time is $k^2 n^{O(\log n)} + n^{O(k)}$. This proves Theorem 5. In this case we need $|\mathbb{F}| > kn^2$.

3. (*Black-box setting*) In this case we only have black-box access to $f = \sum_{i \in [k]} f_i$. Let $f_v(x_1, \ldots, x_n) = f(x_1 + v_1, \ldots, x_n + v_n)$. Then it is easy to see that $f \equiv 0 \iff \forall v \in \mathcal{A}_k, f_v|_{\mathcal{W}_{7k}^n} \equiv 0$. In this case the running-time is $n^{O(\log n + k)}$. This proves Theorem 2. In this case we need $|\mathbb{F}| > kn^4$.

# References

[1] A. Shpilka and I. Volkovich. Read-once polynomial identity testing. In *Proceedings of the 40th Annual STOC*, pages 507–516, 2008.

[2] A. Shpilka and I. Volkovich. Improved polynomial identity testing of read-once formulas. In *Approximation, Randomization and Combinatorial Optimization. Algorithms and Techniques, volume 5687 of LNCS*, pages 700–713, 2009.

[3] M. Agrawal. Proving lower bounds via pseudo-random generators. In *Proc. 25th Annual Conference on Foundations of Software Technology and Theoretical Computer Science*, pages 92–105, 2005.

[4] N. Saxena. Progress of polynomial identity testing. Technical Report ECCC TR09-101, Electronic Colloquium in Computational Complexity, 2009.

[5] J.T. Schwartz. Fast probabilistic algorithms for polynomial identities. *J. Assn. Comp. Mach.*, 27:701–717, 1980.

[6] R. Zippel. Probabilistic algorithms for sparse polynomials. In *Proceedings of the International Symposium on Symbolic and Algebraic Manipulation (EUROSAM '79), volume 72 of Lect. Notes in Comp. Sci.*, pages 216–226. Springer Verlag, 1979.

[7] V. Kabanets and R. Impagliazzo. Derandomizing polynomial identity testing means proving circuit lower bounds. *Computational Complexity*, 13(1–2):1–44, 2004.

[8] M. Ben-Or and P. Tiwari. A deterministic algorithm for sparse multivariate polynomial interpolation. In *Proc. 20th Annual ACM Symposium on the Theory of Computing*, pages 301–309. ACM, 1988.

[9] A.R. Klivans and D.A. Spielman. Randomness efficient identity testing of multivariate polynomials. In *Proc. 33rd Annual ACM Symposium on the Theory of Computing*, pages 216–223, 2001.

[10] R. Lipton and N. Vishnoi. Deterministic identity testing for multivariate polynomials. In *Proceedings of the fourteenth annual ACM-SIAM symposium on Discrete algorithms (SODA 2003)*, pages 756–760, 2003.

[11] Z. Dvir and A. Shpilka. Locally decodable codes with two queries and polynomial identity testing for depth 3 circuits. *SIAM J. Comput.*, 36(5):1404–1434, 2006.

[12] N. Kayal and N. Saxena. Polynomial identity testing for depth 3 circuits. *Computational Complexity*, 16(2):115–138, 2007.

[13] V. Arvind and P. Mukhopadhyay. The monomial ideal membership problem and polynomial identity testing. In *Proceedings of the 18th International Symposium on Algorithms and Computation (ISAAC 2007), volume 4835 of Lecture Notes in Computer Science*, pages 800–811. Springer, 2007.

[14] Z.S. Karnin and A. Shpilka. Deterministic black box polynomial identity testing of depth-3 arithmetic circuits with bounded top fan-in. In *Proc. 23rd Annual IEEE Conference on Computational Complexity*, pages 280–291, 2008.

[15] N. Kayal and S. Saraf. Black box polynomial identity testing of depth-3 circuits. In *Proc. 49th Annual IEEE Symposium on Foundations of Computer Science*, 2009.

[16] Z.S. Karnin, P. Mukhppadhyay, A. Shpilka, and Ilya Volkovich. Deterministic identity testing of depth 4 multilinear circuits with bounded top fan-in. Technical Report TR09–116, Electronic Colloquium on Computational Complexity (ECCC), November 2009.

[17] R. Raz and A. Shpilka. Deterministic polynomial identity testing in non commutative models. *Computational Complexity*, 14(1):1–19, 2005.

[18] M. Agrawal and V. Vinay. Arithmetic circuits: A chasm at depth four. In *Proc. 49th Annual IEEE Symposium on Foundations of Computer Science*, pages 67–75, 2008.

[19] L. Valiant. Completeness classes in algebra. Technical Report CSR-40-79, Dept. of Computer Science, University of Edinburgh, April 1979.

[20] L. Lovász. On determinants, matching, and random algorithms. In *FCT'79: Fundamentals of Computation Theory*, pages 565–574, 1979.

[21] K. Mulmuley, U. Vazirani, and V. Vazirani. Matching is as easy as matrix inversion. *Combinatorica*, 7:105–113, 1987.

[22] E. Allender, K. Reinhardt, and S. Zhou. Isolation, matching and counting uniform and nonuniform upper bounds. *J. Comput. Syst. Sci.*, 59(2):164–181, 1999.

[23] S. Datta, R. Kulkarni, and S. Roy. Deterministically isolating a perfect matching in bipartite planar graphs. In *Proc. 25th Annual Symposium on Theoretical Aspects of Computer Science*, volume 08001 of *Leibniz Int. Proc. in Informatics*, pages 229–240, 2008.

[24] M. Jansen. Weakening assumptions for deterministic subexponential time non-singular matrix completion, 2010. To Appear, 27th International Symposium on Theoretical Aspects of Computer Science (STACS 2010).

[25] N. Alon. Combinatorial nullstellensatz. *Combinatorics, Probability and Computing*, 8(1–2):7–29, 1999.

# A  Figure 3

Figure 3 shows an RO-ABP computing $x_1 x_2 + x_2 x_3 + x_{n-1} x_n$, when $n$ is even. The case when $n$ is odd is dealt with similarly. Unlabeled edges are labeled with 1.

# B  Example : RO-ABPs Are Not Universal

**Proposition 6.** *The degree-2 elementary symmetric polynomial $e_n(x_1, x_2, \ldots, x_n) = \prod_{1 \leq i < j \leq n} x_i x_j$, $n \geq 3$ can not be computed by a RO-ABP.*

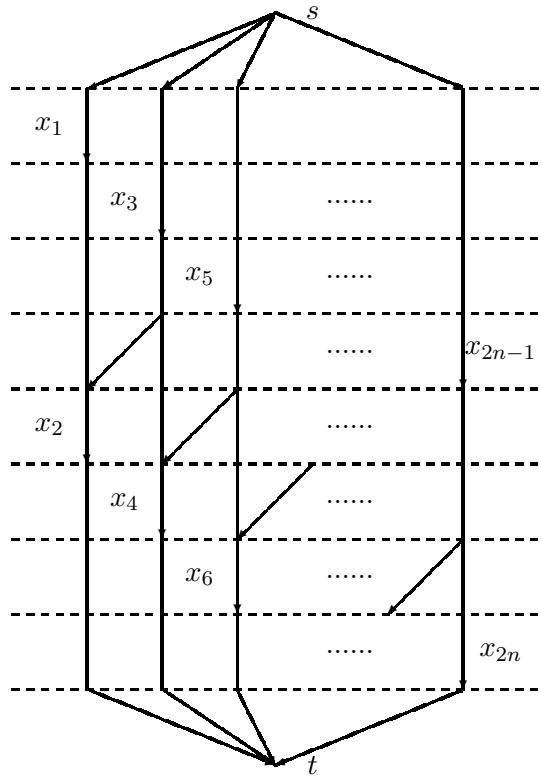

Figure 3: A RO-ABP computing $x_1x_2 + x_2x_3 + \ldots + x_{2n-1}x_{2n}$.

*Proof.* For the purpose of contradiction, suppose that some RO-ABP $A$ computes $e_n$. For any $x_i$ denote the edge it labels by $g_i = (s_i, t_i)$. We can define an ordering $<$ among $g_i$'s, by taking $g_i < g_j$ if and only if the polynomial computed by the subprogram $A(t_i, s_j)$ has a nonzero constant term. Due to the fact that $A$ is a DAG, we have for any $i, j$, if $x_i < x_j$, then not $x_j < x_i$.

The fact that for every $(i, j)$ pair, $x_i x_j$ appears as a term in $e_n$ implies that for any $i \neq j$, we have one of $x_i < x_j$ or $x_j < x_i$. Incidently, note this implies the ordering is transitive. Namely, if $x_i < x_j$ and $x_j < x_k$, then $s_j$ must be reachable from $t_i$, and $s_k$ must be reachable from $t_j$ in $A$, but then $s_i$ can not be reachable from $t_k$. Hence not $x_k < x_j$, which implies $x_j < x_k$.

In any case, observe there is a permutation $\phi : [n] \to [n]$ for which $x_{\phi(1)} < x_{\phi(2)} < \cdots < x_{\phi(n)}$. This implies that $\prod_{i \in [n]} x_i$ appears as a term in the polynomial computed by $A$, which is a contradiction. $\square$

## C  Non-Black-Box Testing a Single RO-ABP

Consider a RO-ABP $A$. Denote the source and sink of $A$ by $s$ and $t$, respectively. Suppose that $x_i$ labels the edge $(s_i, t_i)$. Wlog. assume that the order of variable layers in $A$ is $x_1, x_2, \ldots, x_n$. We have the following easy proposition:

**Proposition 7.** *Suppose $1 \leq i_1 < i_2 < \cdots < i_k \leq n$. For a RO-ABP $A$, $x_{i_1} x_{i_2} \ldots x_{i_k}$ appears in $\hat{A}$ if and only if the constant terms in $\hat{A}(s, s_{i_1})$, $\hat{A}(t_{i_m}, s_{i_{m+1}})$, for all $m \in [k-1]$, and $\hat{A}(t_k, t)$ are not zero.*

We build a directed graph $G_A = (V, E)$ for RO-ABP $A$ with vertex set $V = \{s, t, x_1, x_2, \ldots, x_n\}$. Edges are given as follows:

1. $(s, x_i)$, if the constant term in $\hat{A}(s, s_i)$ is nonzero.

2. $(x_i, t)$, if the constant term in $\hat{A}(t_i, t)$ is nonzero.

3. $(x_i, x_j)$, $i < j$, if the constant term in $\hat{A}(t_i, s_j)$ is nonzero.

We have the following corollary of Proposition 7:

**Corollary 5.** *$\hat{A}(x_1, \ldots, x_n) \equiv 0$ if and only if $t$ is not reachable form $s$ in $G_A$.*

The algorithm for testing $A$ is to construct $G_A$ and to test connectivity. This can be done in time $O(n^2 s)$, where $s$ bounds the size of $A$.